\documentclass[doublecol]{epl2}

\title{Role of activity in human dynamics}
\author{Tao Zhou\inst{1,2} \and Hoang Anh-Tuan Kiet\inst{3} \and
  Beom Jun Kim\inst{3} \and  Bing-Hong Wang\inst{1} \and Petter Holme\inst{4}}
\shortauthor{T. Zhou \textit{et al.}}

\institute{
  \inst{1} Department of Modern Physics, University of Science and
Technology of China -- Anhui, Hefei 230026, P. R. China\\
  \inst{2} Department of Physics, University of Fribourg -- CH--1700,
Fribourg, Switzerland\\
  \inst{3} Department of Physics, BK21 Physics Research Division, and
  Institute of Basic Science, Sungkyunkwan University -- Suwon
  440--746, Republic of Korea\\
  \inst{4} Department of Computational Biology, School of Computer Science
  and Communication -- Royal Institute of Technology, 100 44
  Stockholm, Sweden
}


\abstract{
  The human society is a very complex system; still, there are
several non-trivial, general features. One type of them is the
presence of
  power-law distributed quantities in temporal statistics. In this
  Letter, we focus on the origin of power-laws in rating of movies. We
  present a systematic empirical exploration of the time between two
  consecutive ratings of movies (the \textit{interevent time}). At an
  aggregate level, we find a monotonous relation between the activity
  of individuals and the power-law exponent of the interevent-time
  distribution. At an individual level, we observe a heavy-tailed
  distribution for each user, as well as a negative correlation
  between the activity and the width of the distribution. We support
  these findings by a similar data set from mobile phone text-message
  communication. Our results demonstrate a significant role of the
  activity of individuals on the society-level patterns of human
  behavior. We believe this is a common character in the
  interest-driven human dynamics, corresponding to (but different
  from) the universality classes of task-driven dynamics.
}

\begin{document}

\maketitle

\section{Introduction}

For decades, the social sciences have studied how large-scale
patterns of human activity emerge from the behavior of
individuals~\cite{schelling}. Until a decade ago, data sets were
typically gleaned from questionnaires, observational studies, etc.;
and understandably rather small. Some statistical quantities need
very large statistics to be seen. One such example is power-law
degree distributions. With the development of information (and
database) technology in the last decade, we can now observe
structures that require large data sets. One such recently observed
phenomenon is the power-law distributions of interevent times of
online activity. This feature can be seen both at the level of
populations~\cite{Mainardi2000, Plerou2000, Masoliver2003,
Scalas2004, Kaizoji2004, Scalas2006} and
individuals~\cite{Barabasi2005, Vazquez2006, Dezso2006}, and cannot
be explained by independent, uniformly random, interaction patterns.
Understanding such emerging communication patterns is essential to
be able to predict the impact of new technologies, the spread of
computer viruses~\cite{Balthrop2004,Vazquez2007}, human
travel~\cite{Brockman2006}, etc.

How do power-laws in response, or interevent, times occur? In a
pioneering work, Barab\'asi~\cite{Barabasi2005} proposed a queuing
model as explanation (later solved analytically~\cite{Vazquez2006,
Vazquez2005, Gabrielli2007}). In this model, the power-law
statistics does not come from a power-law distributed trait of the
agents, but emerge from interaction between the agents and the
environment. Barab\'asi's model gives response times of two
universality classes---one with power-law exponent $\alpha=1$
(observed in e-mail communication~\cite{Eckmann2004, Barabasi2005}),
and a class with $\alpha=1.5$ (observed in surface mail
communication~\cite{Oliveira2005}). The behavioral origin of
power-law tails according to Barab\'asi's model~\cite{Barabasi2005},
is that the individuals use a \emph{highest-priority-first} (HPF)
protocol to decide which task needs to be executed first (rather
than a first-in-first-out strategy). However, power-laws have been
observed in systems driven by individuals arguably not guided by
task-lists (e.g., web browsing~\cite{Dezso2006}, networked
games~\cite{Henderson2001} and online chatting~\cite{Dewes2003}). In
this work, we perform a detailed study of such a system, namely an
online infrastructure for rating movies. Our primary quantity is the
time $\tau$ between two consecutive movie ratings. The distribution
$p(\tau)$ of the aggregated data follows a power law spanning more
than two orders of magnitude. More interestingly, we observe a
monotonous relation between the power-law exponent and the mean
activity in the group (see below how to divide the whole population
into several groups). This suggests that the activity of individuals
is one of the key ingredients determining the distribution of
interevent times.

\begin{figure}\center
  \scalebox{0.7}[0.7]{\includegraphics{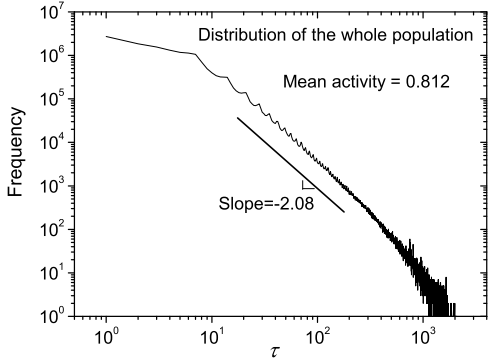}}
  \caption{The distribution of interevent time in the population
    level, indicating that $p(\tau)\sim \tau^{-2.08}$. The solid line
    in the log-log plot has slope $-2.08$. The data exhibits weekly
    oscillations, reflecting a weekly periodicity of human
    behavior, which has also been observed in e-mail
    communication~\cite{Holme2003}.
  }\label{fig:1}
\end{figure}

\section{Data source}

Our data source, obtained from www.netflixprize.com, is collected by
a large American company for mail order DVD-rentals, Netflix. The
users can rate movies online. This information is used to give the
users personalized recommendations. The data was made public as a
part of a competition for the better recommender system. In total,
the data comprises $M=17{,}770$ movies, $N=447{,}139$ users and
$\sim 9.67\times 10^7$ records. Each record consists of four
elements: a user ID $i$, a movie ID $\alpha$, the user's rating
(from $1$ to $5$) $v_{i\alpha}$, and the time of the rating
$t_{i\alpha}$). Tracking the records of a given user $i$, one can
get $k_i-1$ interevent times where $k_i$ is the number of movies $i$
has already seen. The time resolution of the data is one day.

\begin{figure}\center
\scalebox{0.6}[0.6]{\includegraphics{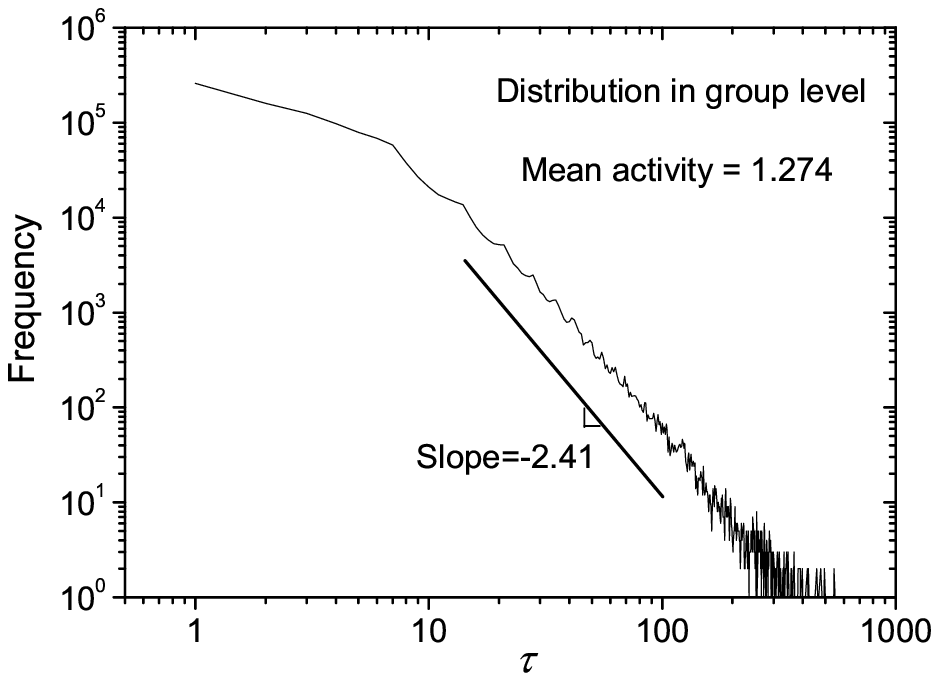}}
\scalebox{0.6}[0.6]{\includegraphics{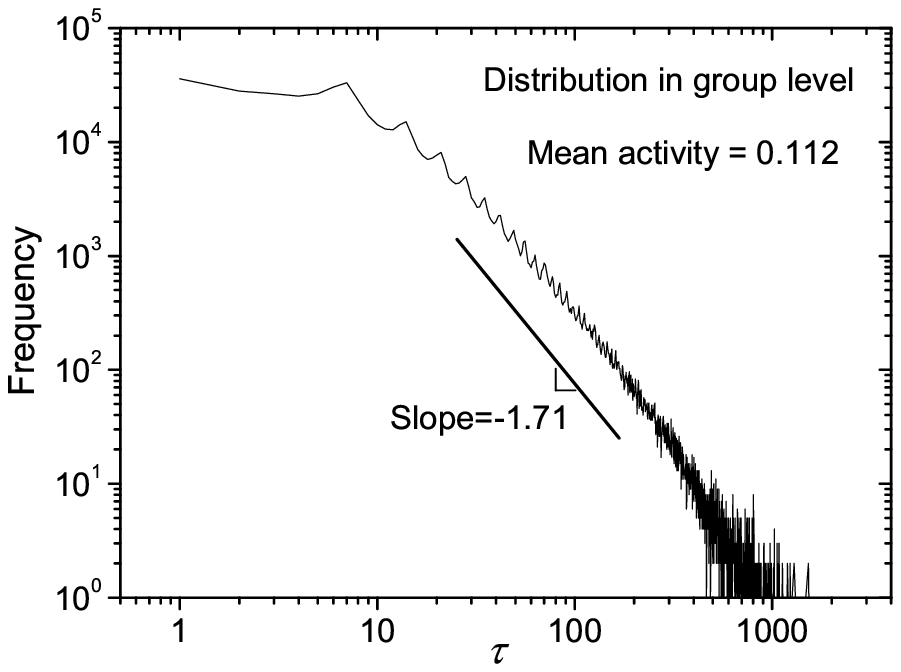}} \caption{The typical
distributions of interevent times at a group level---group 4 (upper
panel) and group 17 (lower panel). The solid lines in the log-log
plot have slopes $-2.41$ and $-1.71$, respectively. The
corresponding mean activities are $1.274$ and $0.112$.}\label{fig:2}
\end{figure}

\section{Interevent time distribution for the whole population}

In Fig.~\ref{fig:1}, we report the interevent time distribution
based on the aggregated data of all users. The distribution follows
a power law, $p(\tau)\sim \tau^{-\gamma}$, for more than two orders
of magnitude. The power-law exponent, $\gamma\approx 2.08$, is
obtained by maximum likelihood estimation~\cite{Goldstein2004}. All
the power-law exponents reported in this Letter are obtained by this
method. To avoid bias from the mentioned oscillation effect, at the
whole-population level, we only include the data points separated by
one week. That is to say, in the calculation of the power-law
exponent, only the data points $F(7), F(14), F(21), \cdots$ are
considered, where $F(\tau)$ denotes the frequency of interevent time
$\tau$. A proposed mechanism for the emergence of power-law
distributions with $\gamma\approx 2.0$ is aggregation of Poissonian
distributions with different, uniformly distributed, characteristic
times~\cite{Hidalgo2006}. However, as we will see later, the
empirical statistics and analysis at group and individual levels
demonstrate that this scaling law cannot be caused by a combination
of Poissonian agents.

\begin{figure}\center
  \scalebox{0.7}[0.7]{\includegraphics{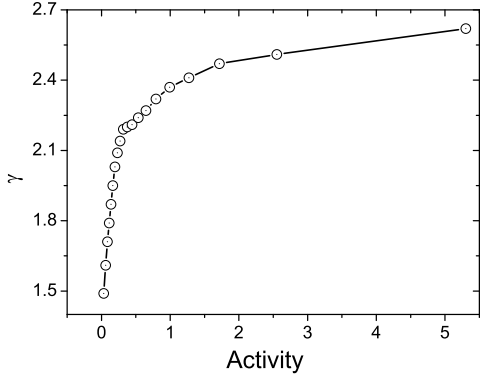}}
  \caption{The relation between power-law exponent $\gamma$ of
    interevent time distribution and mean activity of each
    group. Each point corresponds to one group. All the exponents are
    obtained by using maximum likelihood estimation and pass the
    Kolmogorov--Smirnov test with threshold quantile
    $0.9$~\cite{Goldstein2004}.
  }\label{fig:3}
\end{figure}

\section{Interevent time distribution for groups}

The HPF protocol~\cite{Barabasi2005} explains heavy tails in
response times of human communication. Nevertheless, we lack an
in-depth understanding of the interevent time distribution in data
sets such as ours. We can probably not explain the aggregated
distribution by identical behavior. A heavy smoker, consuming fifty
cigarettes per day, would not make a long pause. Events separated by
longer times would (assuming smoking patterns follows the same
statistics) come from other people---occasional party-smokers,
mischievous adolescents, or similar. Similarly, the other end of the
spectrum in Fig.~\ref{fig:1} probably corresponds to other persons.
To get at this we measure the \emph{activity}
$A_i$~\cite{Ghoshal2006}---the frequency of events of an individual:
$A_i=n_i/T_i$, where $n_i$ is the total number of records of $i$,
and $T_i$ is the time between the first and the last event of $i$.
In other words, $A_i$ is the frequency of movie ratings of $i$. As
shown in Fig.~\ref{fig:1}, the mean activity, averaged over all
users, is $\langle A \rangle=0.812$.

\begin{figure}\center
\scalebox{0.8}[0.8]{\includegraphics{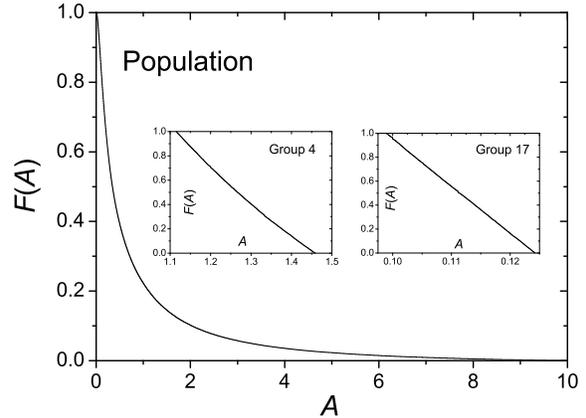}}
\caption{Cumulative distribution of activities for all the
  individuals. The distribution is intermediate between exponential
  and power-law. The insets display the same measure for group 4 and
  group 17, respectively.
}\label{fig:4}
\end{figure}

To investigate the role of activity, we sort the users by activity
in a descending order, and then divide this list into twenty groups,
each of which has almost the same number of users. Accordingly, the
mean activity of each group obeys the inequality $\langle A
\rangle_1>\langle A \rangle_2>\cdots>\langle A \rangle_{20}$. In
Fig.~\ref{fig:2}, we report two typical distributions of interevent
time at a group level. Both these distributions follow power-laws.
Note that the group with lower activity has power-law exponent,
giving a longer average interevent time. The corresponding
distributions for the other groups follow power-law forms as well,
but with different exponents. In Figure~\ref{fig:3} we diagram the
exponent as a function of activity. There is a non-trivial,
monotonous increase of the exponent with the activity. This
relation, in accordance with our smoker example above, indicates the
significant role of activity for the observed, aggregate behavior.
Note that, for a mathematically ideal power-law distribution
$p(\tau)\sim \tau^{-\gamma}$, the exponent $\gamma$ has a one-to-one
correspondence with $A$ from the relation
\begin{equation}
\gamma(A)=1+\frac{1}{1-A} ~ , ~ 0<A<1~.
\end{equation}
For $A>1$, there is no corresponding normalized probability
distribution, of $\tau$, of a power-law form. However, the situation
in the real data is very different. As shown in Figs.~\ref{fig:1} and
\ref{fig:2}, the activity are mainly determined by the drooping head
of $p(\tau)$, not the tail used to calculate $\gamma$ (we consider
$\tau=7,14,21,\cdots$ only). A similar case can be found
in~\cite{Barabasi2005} and its supplementaries, where a peak at
$p(\tau=1)$, which was ignored in the calculation of $\gamma$, mainly
describes the individual activity.

\begin{figure}\center
\scalebox{0.42}[0.42]{\includegraphics{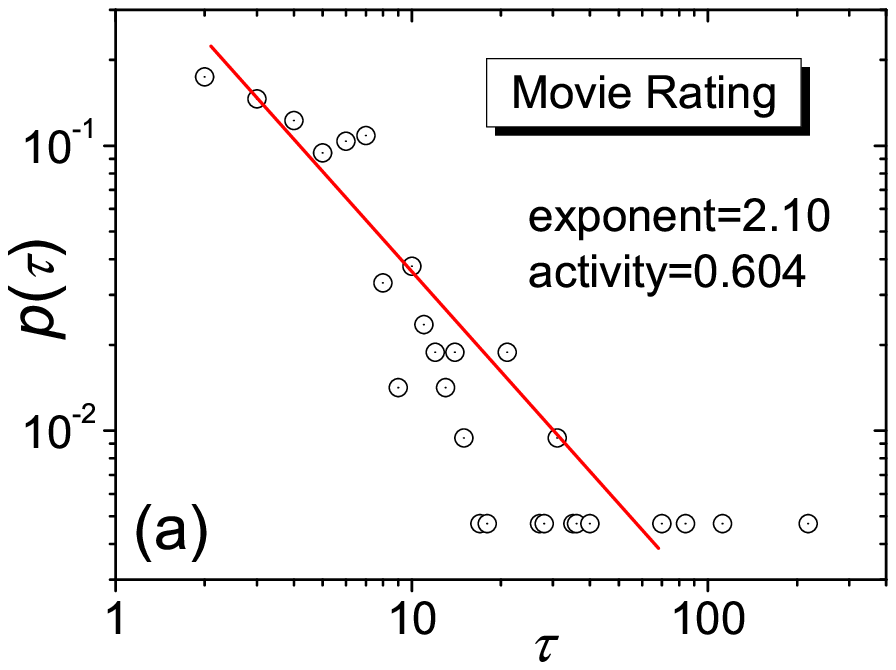}}
\scalebox{0.41}[0.41]{\includegraphics{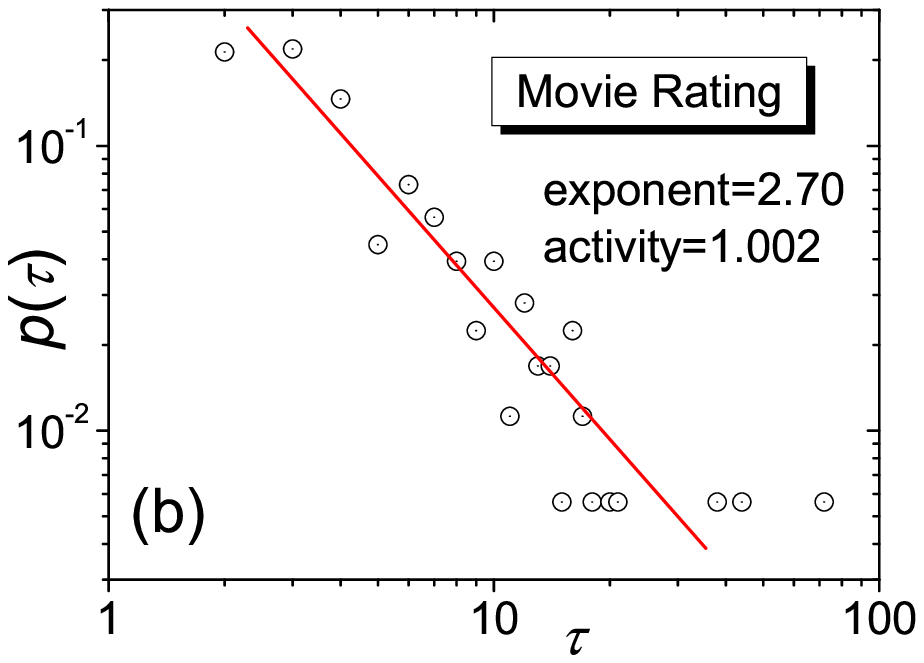}}
\scalebox{0.39}[0.39]{\includegraphics{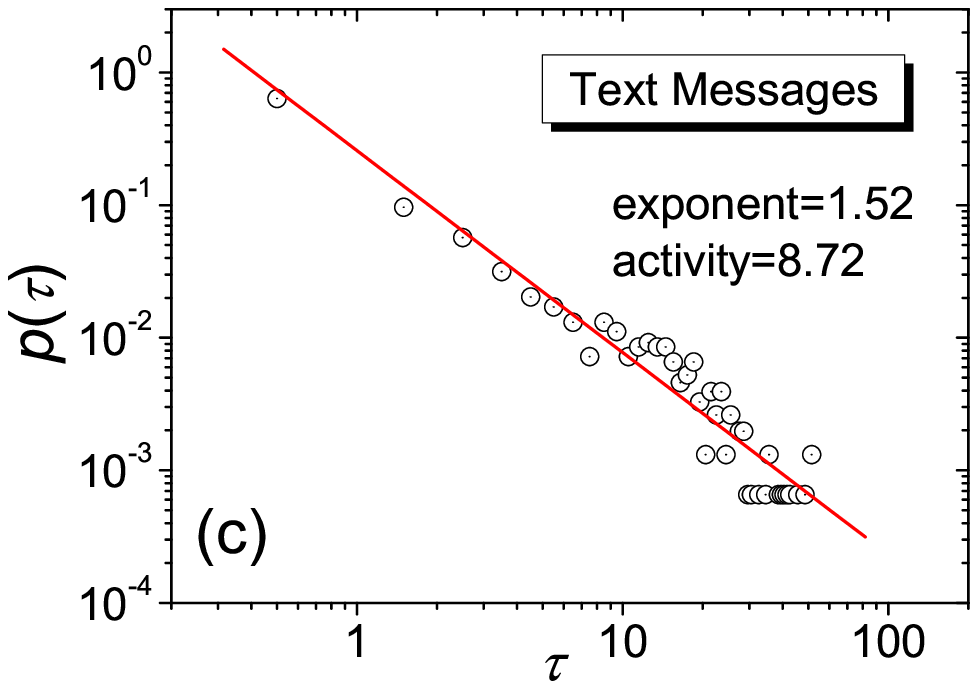}}
\scalebox{0.39}[0.39]{\includegraphics{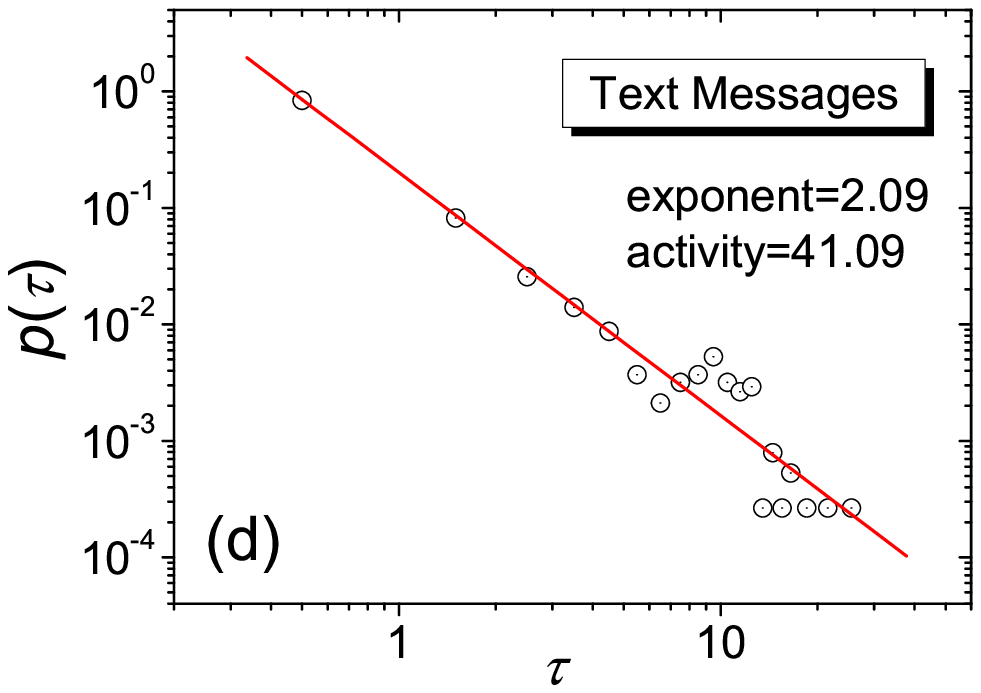}} \caption{(Color
online) The interevent time distribution between, (a)--(b) two
consecutive movie ratings by two Netflix users, and (c)--(d) two
consecutive sending of text-messages by two mobile telephone users.
The time unit for (a) and (b) is one day, and for (c) and (d) one
hour. Under the threshold quantile $0.9$, distributions in (a) and
(b) can not pass the Kolmogorov--Smirnov test, while the (c) and (d)
do pass it. }\label{fig:5}
\end{figure}

If every monitored individual has a Poisson distributed activity at
separate rate $A$, then the distribution of interevent time should
be~\cite{Hidalgo2006}
\begin{equation}
p(\tau)\sim f(A)\tau^{-2}~,
\end{equation}
where $f(A)$ is the activity distribution of individuals. Since the
power-law exponent in population level is close to 2, if it results
from an aggregation of Poissonian individuals, the activity
distribution should follow a uniform pattern. However, as shown in
the main plot of Fig.~\ref{fig:4}, the activity distribution in
population level is not uniform. In contrast, as reported in the
insets of Fig.~\ref{fig:4}, the cumulative distribution $F(A)$ for
group 4 and group 17 can be well fitted by a straight line,
suggesting a uniform distribution $f(A)$, while the exponents
$\gamma_4$ and $\gamma_{17}$ are far from each other, and both
different from 2. Therefore, the heavy-tailed nature at the group
level cannot originate from homogeneous Poissonian individuals. To
our knowledge, it is the first time one has observed, a monotonous
relation between power-law exponent of interevent time distribution
and a certain measure (i.e.\ activity). We believe this analysis
illustrate the important role of the individual activity in the
aggregate pattern of human behavior.

\begin{figure}\center
\scalebox{0.7}[0.7]{\includegraphics{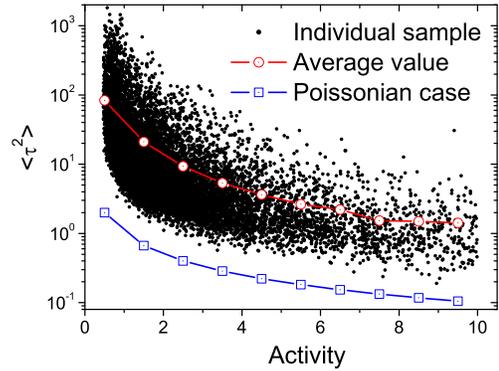}} \caption{(Color
online) Scatter plot showing the second moment $\langle
\tau^2\rangle$ and activity, indicating a negative correlation. The
red curve shows the average value of $\langle \tau^2\rangle$ for a
given activity, and the blue curve represents the case of Poisson
distribution whose expected value is given as the inverse of
activity.
}\label{fig:6}
\end{figure}

\section{Interevent time distribution for individuals}

To continue tying together micro- and macro phenomena, we look closer
at the behavior of individual agents. In particular, we investigate
whether or not the monotonous relation between activity and power-law
exponent also holds at an individual level.

Figs.~\ref{fig:5}(a) and (b) report the interevent time distribution
$p(\tau)$ of two individual users. We observe a similar relation as
for the group level statistics. That is to say, the less active
agent has a broader distribution and smaller power-law exponent.
Although the distributions shown in Figs.~\ref{fig:5}(a) and (b)
show heavy-tailed forms, they do not pass the Kolmogorov--Smirnov
test with threshold quantile 0.9~\cite{Goldstein2004}. We believe
this can be explained by the relative short sample times of the
individual records. (The typical duration of individual records, in
our case, range from a few months to a few years. This range is not
as impressive as, e.g.\ Refs.~\cite{Oliveira2005,Vazquez2007b} where
surface mail is studied for a period of more than half century with
a resolution in days.) It may be the case that a credible power-law
scaling will emerge after a sufficient while; however, so far, we
cannot claim that typical $\tau$-distributions follow power-law
forms. Nevertheless, almost every user has a heavy-tailed
distribution (that is, much broader than a Poisson distribution with
the same average interevent time $\langle \tau\rangle$). We use the
second moment, $\langle \tau^2\rangle=\int
\tau^2p(\tau)\,\mathrm{d}\tau$, to measure the width of $p(\tau)$.
As seen in Fig.~\ref{fig:6}, all individual distributions have much
larger $\langle \tau^2\rangle$ than the Poisson distributions with
the same $\langle \tau\rangle$. Moreover, we observe a negative
correlation between $\langle \tau^2\rangle$ and $A$, which can be
seen as an individual-level variant of the relation in
Fig.~\ref{fig:3}. Although the negative correlation can also be
detected in Poisson distributions, this finding is interesting since
it highlights the activity, as opposed to universality classes, as a
signifier of human dynamics.

To check the generality of our observations of the relation between
activity and interevent time patterns, we investigate another
empirical data set of mobile phone text-message
communication. The data set comprise all messages sent and received by
20 users over half a year. Figure~\ref{fig:5}(c) and (d) report two
typical interevent time distributions. These show yet more credible
power-laws than those in the Netflix data (Fig.~\ref{fig:5}(a) and (b)).
Actually, in this data set, all users show a power-law
distribution passing the Kolmogorov--Smirnov test. (Note that, the
time resolution of the text-message data is seconds. Thus, half a
year is long compared to the Netflix data.) The
activities and exponents belong to the intervals $A\in [6.09, 60.72]$
and $\gamma \in [1.41,2.25]$. Even at the individual level (which is
sensitive to fluctuations in personal habits), an almost monotonous
relation between $A$ and $\gamma$ is observed (with the exception of
two users that show a slight deviation). A similar relation can also
be found in data of online Go (duiyi.sports.tom.com); in this data the
individual records span years, and the resolution is hours). Here, the
more active players also have larger power-law exponents and
narrower interevent time distributions. However, for commercial reasons,
the aggregated data cannot be freely downloaded. Therefore, for the
text-message and online Go data we cannot analyze the aggregate level
statistics.

\section{Conclusions}

In previous works, the heavy-tailed interevent time distribution has
been explained by a queuing mechanism in the decision making of
agents. This is a relevant scenario for task-driven situations (such
as e-mail~\cite{Barabasi2005} or surface mail~\cite{Oliveira2005}
communication). However, similar, heavy-tailed distributions also
exists in many interest-driven systems (e.g.\ web
browsing~\cite{Dezso2006}, networked computer games~\cite{Henderson2001}, online
chat~\cite{Dewes2003}; or, as our examples, text-message sending, and
movie rating), where no tasks are waiting to be executed. As opposed
to focusing on universality classes (as for task-driven systems), we
highlight a common character in interest-driven systems: the power-law
exponents are variable in a wide range with a strongly positive
correlation to the individual's activity. This finding is helpful for
further understanding the underlying origins of heavy tails of
interest-driven systems. A power-law distribution of activity, might
also be a factor in the dynamics of task-driven systems. This is
reminiscent of the power-law distribution of extinction events (that
can be explained by both the internal dynamics of evolution, and a
power-law distribution of the magnitudes of natural
disasters~\cite{soc}).

\acknowledgments
We thank Mr.\ Wei Hong for providing us the text-message
data. B.J.K. acknowledges support from the Korea Science and
Engineering Foundation by the grant No.\ R01--2007--000--20084--0, and
H.A.T.K. acknowledges support from the Korea Research Foundation with
grant No.\ KRF--2006--211--C00010.  B.H.W. acknowledges the 973 Project
2006CB705500, T.Z. was supported by NNSFC--10635040,
P.H. acknowledges support from The Swedish Foundation for Strategic
Research.

\end{document}